\documentclass[5p,sort&compress]{elsarticle}

\usepackage{graphicx}
\usepackage[colorlinks=false]{hyperref}
\usepackage{array}
\usepackage{dcolumn}
\usepackage{lineno}

\modulolinenumbers[5]

\usepackage{amsmath}
\usepackage[hypcap,small]{caption}

\newcolumntype{d}[1]{D{.}{.}{#1}}

\begin{document}

\begin{frontmatter}

\title{Space-charge effects in high-energy photoemission \tnoteref{t1}}
\tnotetext[t1]{\textcopyright{} 2016. This manuscript version is made available under the CC-BY-NC-ND 4.0 license\\ \url{http://creativecommons.org/licenses/by-nc-nd/4.0/}.}

\date{\today}

\author[scienze_roma3,cnism]{Adriano Verna\corref{cor}}
\ead{adriano.verna@uniroma3.it}
\cortext[cor]{Corresponding author}

\author[scienze_roma3]{Giorgia Greco}

\author[scienze_roma3,dott_roma3]{Valerio Lollobrigida}

\author[scienze_roma3,cnism]{Francesco Offi}

\author[scienze_roma3,cnism]{Giovanni Stefani}

\address[scienze_roma3]{Dipartimento di Scienze, Universit\`{a} degli Studi Roma Tre, Via della Vasca Navale 84, I-00146 Roma, Italy.}

\address[dott_roma3]{Scuola Dottorale in Matematica e Fisica, Universit\`{a} Roma Tre, Via della Vasca Navale~84, I-00146 Roma, Italy.}

\address[cnism]{CNISM Unit\`{a} di Roma Tre, Via della Vasca Navale 84, I-00146 Roma, Italy.}

\begin{abstract}
Pump-and-probe photoelectron spectroscopy (PES) with femtosecond pulsed sources opens new perspectives in the investigation of the ultrafast dynamics of physical and chemical processes at the surfaces and interfaces of solids. Nevertheless, for very intense photon pulses a large number of photoelectrons are simultaneously emitted and their mutual Coulomb repulsion is sufficiently strong to significantly modify their trajectory and kinetic energy. This phenomenon, referred as space-charge effect, determines a broadening and shift in energy for the typical PES structures and a dramatic loss of energy resolution. In this article we examine the effects of space charge in PES with a particular focus on time-resolved hard X-ray ($\sim$10~keV) experiments. The trajectory of the electrons photoemitted from pure Cu in a hard X-ray PES experiment has been reproduced through $N$-body simulations and the broadening of the photoemission core-level peaks has been monitored as a function of various parameters (photons per pulse, linear dimension of the photon spot, photon energy). The  energy broadening results directly proportional to the number $N$ of electrons emitted per pulse (mainly represented by secondary electrons) and inversely proportional to the linear dimension $a$ of the photon spot on the sample surface, in agreement with the literature data about ultraviolet and soft X-ray  experiments. The evolution in time of the energy broadening during the flight of the photoelectrons is also studied. Despite its detrimental consequences on the energy spectra, we found that space charge has negligible effects on the momentum distribution of photoelectrons and a momentum broadening is not expected to affect angle-resolved experiments. Strategy to reduce the energy broadening and the feasibility of hard X-ray PES experiments at the new free-electron laser facilities are discussed.
\end{abstract}



\end{frontmatter}



\section{Introduction}

Photoelectron spectroscopy (PES) is one of the most powerful tool for the study of the electronic and chemical properties of materials \cite{Fadley_10,Hufner_book_03,Mariani_incoll_15}. Photoemission from core levels gives information about the chemical composition of a surface and the oxidation states of the various elements whereas electrons emitted from valence bands are exploited to study the density of states and the characteristics of the continuous bands close to the Fermi level. Angle-resolved photoemission spectroscopy (ARPES) is ba\-sed on the detection of energy and momentum of photoemitted electrons and provides a very detailed structure of the valence bands of a material \cite{Kevan_book_92,Damascelli_04}. ARPES is nowadays a fundamental instrument to  access the electronic properties of solids and proved itself irreplaceable in the study of  correlated electrons in complex oxides \cite{Damascelli_03}. Ultraviolet (UV) lamps (15--20~eV) and soft X-ray tubes (Mg K$_\alpha$~1253.6~eV, Al K$_\alpha$~1486.6~eV) have been traditionally used to produce monochromatic exciting radiation and the advent of synchrotron radiation in the last decades has provided a new high-brilliance and high-intensity source with the fundamental advantage of a continuous tunability of the photon energy. In most experiments the detected photoelectrons have a kinetic energy in the 15--2000~eV range. For the relatively short mean free path of the electrons in this energy range and for the mild dependence of this quantity on the properties of the material, the probing depth is approximately between 0.5 and 2~nm from the top of the sample surface. Strong sensitivity to surface properties is one of the most prominent features of PES and has determined its development as a pivotal experimental techniques in the broad field of surface science \cite{Powell_09}. Nevertheless, in recent years the use of hard X-rays (5--10~keV) produced at synchrotron radiation facilities has received a considerable interest in PES experiments which detect high-energy elastic photoelectrons originating from a depth of 10~nm or more below the surface \cite{Sacchi_05}. Hard X-ray photoemission spectroscopy (HAXPES) provides information on the bulk properties of the examined material with an energy resolution comparable to that of lower-energy photoemission experiments and opens new perspectives in the study of buried interfaces, usually unapproachable with traditional ultraviolet and soft X-ray PES \cite{Panaccione_12,Borgatti_13}. The use of hard X-rays in ARPES proved to be effective in obtaining the momentum-dispersion of valence bands with bulk sensitivity \cite{Gray_AX_11}.

The extension of the well-established PES techniques to time-resolved investigations constituted in the last years one of the most promising developments of this family of experimental tools \cite{Bokor_89,Stolow_04,Fadley_10,Schonhense_15}. In pump-and-probe experiments, a pump pulse is used to perturb the examined system at an instant $t=0$ and a second radiation pulse, impinging on the sample after a time $\Delta t$, is used to probe the properties of the excited system. Examples of phenomena investigated through time-resolved photoelectron spectroscopy (tr-PES) are the dynamics of surface reactions \cite{DellAngela_13}, the melting of a charge density waves \cite{Schmitt_08}, the ultrafast demagnetization of ferromagnetic thin films \cite{Rhie_03,Cinchetti_06}, charge dynamics in metals \cite{Cavalieri_07}, image potential states \cite{Pickel_06}. Time-resolved ARPES has also been extensively used, for example to investigate the dynamics of the long-range charge order in solids \cite{Rohwer_11}. The duration of the pump and probe pulses must be much shorter than the time resolution requested by the experiment. The performances of tr-PES are strictly connected with the development of pulsed radiation sources capable to produce a huge number of highly collimated photons concentrated in picosecond or sub-picosecond pulses. Ti:sapphire laser is often employed as pulsed ultra-violet probe. Its fundamental frequency (1.5~eV) is multiplied making use of  high harmonic generation (HHG) in gases \cite{Cavalieri_07,Rohwer_11} or non-linear crystals \cite{Schmitt_08,Cinchetti_06}; the infra-red fundamental harmonic of the laser is generally used as the pump radiation. At higher photon energies (above $\sim$100~eV) the pulsed structure of synchrotron radiation has been exploited in third-generation facilities for tr-PES experiments with a time resolution of tens of picoseconds \cite{Glover_03,Pietzsch_07}. In this case the pump radiation is provided by a pulsed laser synchronized with the synchrotron's storage ring. In order to monitor processes occurring on the scale of femtoseconds with X-ray PES a probe source with comparable pulse duration must be implemented. The use of HHG in particular conditions \cite{Popmintchev_12} and the femtosecond slicing of synchrotron radiation \cite{Schoenlein_00,Khan_S_06} are proposed as possible candidates but they still generate a photon flux too weak to acquire photoelectron spectra in an efficient way. Enormous efforts were focused in recent years in developing the novel free-electron laser (FEL) facilities which provide soft and hard X-ray pulses of extreme short duration (1--100~fs) and unequaled intensity and brilliance and opened perspectives to a very wide range of new experiments \cite{Popmintchev_12,RebernikRibic_12}. Pioneering works demonstrated the feasibility of tr-PES with FEL sources \cite{Pietzsch_08,Hellmann_12a,Oura_14} and pump-and-probe experiments have already given important results about the ultrafast dynamics of chemical and physical processes \cite{DellAngela_13,Hellmann_10}. The development of tr-PES with FEL sources has increased the demand for new, high-resolution and ultrafast electron spectrometers. The excitation by radiation pulses regularly spaced in time makes a time-of-flight (TOF) system the best suited and most natural choice in designing an experimental apparatus \cite{Ovsyannikov_13, Lollobrigida_15}.

One of the fundamental problems with tr-PES is that very intense and very short radiation pulses produce a huge number of photoelectrons that are emitted at the same time from the sample surface and interact with each other in vacuum.  A photoemission spectrum reports the number of detected photoelectrons as a function of their kinetic energy. Most of the information is carried by the small fraction of the primary photoelectrons that, from inside the material, reach the surface without experiencing an inelastic scattering process with other electrons or plasmons and overcome the surface potential barrier. These elastic electrons originate the principal structures in a photoemission spectrum: continuous bands near the Fermi level, core-level peaks and Auger peaks. However, when a sufficient number of electrons simultaneously leave the surface their Coulombic interaction causes a variation in their kinetic energy and momentum that affects the features of the recorded energy spectrum inducing a broadening of the PES structures, with a consequent loss in energy resolution.

The excess electric charge due to electrons and ions dispersed in vacuum (or in a dielectric where screening processes are prevented) is known as space charge and the results of the mutual interactions among particles are denoted space-charge effects. These effects must be taken in very careful account in designing devices operating with electron beams like cathode tubes, electron guns and electron microscopes \cite{Kruit_incoll_97}. The most usual effect of the mutual interaction of electrons in vacuum is the reduction of the thermionic and photoemission currents known as the space-charge limit (SCL), which is fundamental in the operation of the vacuum tubes \cite{Belinov_09}. Space-charge effects in photoemission were already well known since the eighties in the first tr-PES experiments using ultraviolet pulsed lasers \cite{Bokor_85}. First systematic theoretical and experimental studies followed \cite{Clauberg_89,Gilton_90} and they evidenced a distortion and broadening of the valence band structures, which increase linearly with the number of photoelectrons per pulses, as well as a reduction of the photoemission current due to the SCL. While Gilton et al. \cite{Gilton_90} presented the first computer simulations of space-charge effects based on the approximation of a continuous charge distribution, in 1996 Long, Itchkawitz and Kabler proposed an oversimplified model to describe the energy broadening $\Delta E$ of the photoemission structures which was predicted to depend roughly on the ratio between the number $N$ of photoelectrons per pulse and the linear dimension $a$ of the radiation spot on the sample surface \cite{Long_96}. Surprisingly, despite its very strong approximations this model was effective in describing the order of magnitude of the energy broadening observed in various tr-PES experiments reported in literature. In 2005 Zhou et al. presented a very systematic experimental study of space-charge effects in photoemission using synchrotron radiation \cite{Zhou_05}. They observed and characterized the energy shift and broadening of Au valence band structures in photoelectron spectra obtained with $h\nu$=35~eV radiation pulses having a duration of $\sim$60~ps and separated by 2~ns (500~MHz) as a function of various parameters (photon flux and photoelectron current, spot dimension, detection angle etc.) They pointed out the fundamental role played by the secondary electrons which have an energy distribution peaked around few eV and constitute the largest part of the overall photoelectrons. They are principally responsible for the broadening of higher-energy photoelectron structures (core-level peaks and valence band features) and their shift towards higher kinetic energies. Due to Coulomb repulsion, fast photoelectrons tend to be pushed forward by slow secondary electrons behind them, whereas slower electrons tend to be retarded or even forced back to the sample. More recently Hellmann et al. \cite{Hellmann_09} presented a detailed theoretical study simulating the space-charge effects on PES structures for very different relevant parameters (number of total photoemitted electrons per pulse, electron kinetic energy of the investigated structure, pulse duration and spot dimension) obtaining a good agreement with the data reported in literature. The same group also investigated the shift and broadening of PES structures when the spot size is reduced to sub-micrometer dimensions for experiments which require high spatial and time resolution \cite{Hellmann_12}. Sch\"onhense~et~al. very recently presented a detailed simulation work about the effects of mutual electron interactions in hard X-ray momentum microscopy \cite{Schonhense_proof}. In recent years many articles reported and discussed experimental evidence of space-charge effects in tr-PES with femtosecond ultra-violet lasers \cite{Passlack_06,Buckanie_09,Graf_10,Leuenberger_11,Faure_12} and free-electron lasers \cite{Pietzsch_08,Hellmann_12a,Oura_14,Oloff_14,DellAngela_15}. In particular, Oloff~et~al. \cite{Oloff_14}  and Dell'Angela~et~al. \cite{DellAngela_15}  investigated in detail the space-charge effects introduced by the photoelectrons emitted by the pump pulse on the PES spectrum produced by the probe radiation and studied their dependence as a function of the delay time between pump and probe pulses.

A phenomenon strictly related to the space charge is the image-charge effect. In presence of a external point charge $q$ the electrons on a metal surface redistribute in order to screen the electric field inside the metal. For a flat metal surface, the potential and the electric field generated by this redistribution is equivalent to those generated by a virtual charge $-q$ symmetric to the real external charge $q$ with respect to the metal plane \cite{Griffiths_book_99}. This virtual charge $-q$ is called mirror or image charge. Zhou et al. \cite{Zhou_05} and Hellmann~et~al. \cite{Hellmann_09} in their simulations included the attractive Coulomb interaction exerted on a photoelectron by the mirror charges of the other electrons of the clouds. The interaction between a photoelectron and its own image is present also in low-rate photoemission and should be included in the work function $\phi_W$ of the metal. Nevertheless, the image-charge method is valid in a static situation and it is not clear if it is applicable with photoelectrons, as already pointed out \cite{Clauberg_89,Zhou_05}. An electron in vacuum with an energy of 10~eV has a velocity of 1.88$\cdot 10^{6}$~m/s, which is comparable with the Fermi velocity of most metals \cite{Ashcroft_book_76}. So it is not sure whether conduction electrons on metal surface can move sufficiently fast to screen instantaneously the photoelectrons. As pointed out by Clauberg and Blacha \cite{Clauberg_89} in any case the holes left behind should exert a Coulomb attraction on the photoelectrons. Simulations by Zhou et al. suggest that the mirror charge significantly reduces the shift towards higher kinetic energy of the photoemission peaks but plays a less important role in the energy broadening.

The energy shift and broadening of the photoemission structures induced by the space charge must be carefully taken into account in  designing  new tr-PES experiments with femtosecond ultra-violet lasers and FELs. However, while the energy shift can be  corrected in the post-mea\-sure\-ment analysis, the energy broadening of PES structure leads to an irrecoverable loss of resolution and information. A reduction of the available photon flux or the enlargement of the area of the beam spot on the sample surface may be required in order to make the space-charge effects negligible or at least tolerable, i.e. lower than the required energy resolution in terms of electronvolts. But this reflects in depleted count rate and spatial resolution. A feasibility study for a new tr-PES apparatus, in particular one designed for a beamline at the emerging FEL facilities, has to carefully consider these limitations \cite{Greco_unpublished}. In this paper we present some ideas concerning the estimation of energy broadening induced by the space charge in photoemission, with a focus on hard X-ray experiments. In particular, following the work by Hellman et al. \cite{Hellmann_09}, we performed for the first time N-body simulations of high-energy (about 10~keV) core-level photoelectrons interacting with low energy secondary electrons using the software Treecode written by Barnes and Hut \cite{Barnes_86,Treecode}.

The paper is structured as follows. In Section~\ref{sec:LIK} we describe the Long, Itchkawitz and Kabler (LIK) model and compare its prediction with data reported in recent literature. We show for the first time that from the LIK model it is possible to estimate the characteristic time $\tau_C$ in which the energy broadening process is accomplished after the exciting pulse.  In Section~\ref{sec:simulation} we briefly describe the Treecode software and its use in the simulation of interacting electrons. In Section~\ref{sec:mono} we present a series of tests with a cloud of mono-energetic electrons emerging from a surface and we show that they verify the predictions of LIK model describing the dependence of energy broadening $\Delta E$ on the number of photoelectrons $N$ and the spot size $a$. Finally in Section~\ref{sec:Cu} we present the simulation of a real case: we consider photoemission from a Cu sample with 8--9~keV photons and we study the shift and broadening of 2p peaks, mostly induced by the low-energy secondary electrons.  The influence of the space charge on the momentum of individual photoelectrons is also considered in view of ARPES applications. In Section~\ref{sec:discussion} we discuss the strategies to minimize the space-charge effects. We also give an estimation of the expected photoelectron count rate when the incident photon flux is sufficiently reduced to have an energy broadening lower than 0.1~eV and consider the feasibility of this type of experiment in a new FEL facility like the European X-FEL at Hamburg \cite{Altarelli_techrep_07}.

\section{The Long-Itchkawitz-Kabler model} \label{sec:LIK}

\begin{figure}
  \centering
  \includegraphics[width=7 cm]{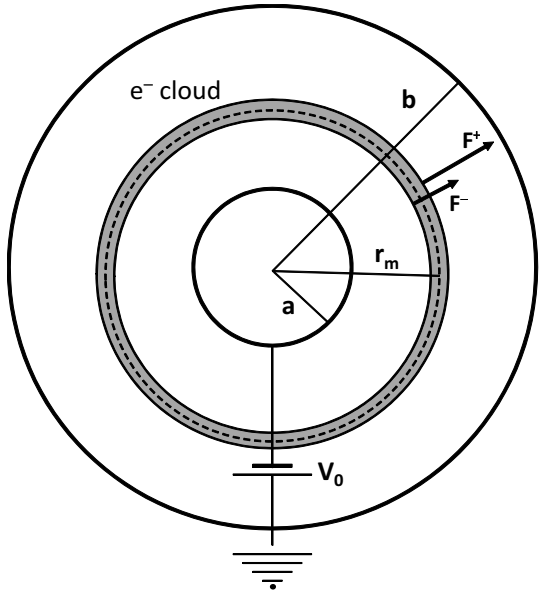}\\
  \caption{An illustration of the simplified model to describe the energy broadening of photoelectrons as described in the paper by Long, Itchkawitz and Keibler \cite{Long_96}.}
  \label{fig:LIK_model}
\end{figure}

Following a suggestion by Clauberg and Blacha \cite{Clauberg_89}, Long, Itchkawitz and Kabler proposed in an appendix of their 1996 paper devoted to tr-PES with synchrotron radiation  an oversimplified model to estimate the energy broadening due to the space charge in photoemission \cite{Long_96}. Their scheme is based on a spherical capacitor and is reproduced in Fig.~\ref{fig:LIK_model}. The ``sample'' is composed by an internal metallic sphere of radius $a$ while an external sphere of radius $b$ acts as a ``detector''. A battery keeps a constant voltage $V_0$ between the two conductors. All the photoelectrons are uniformly emitted from the internal sphere perpendicularly to  its surface with the same initial kinetic energy $E_0$. The time interval in which electrons are emitted (i.e. the photon pulse duration $\tau_p$) is much shorter than the time of flight between the two spheres. The electron cloud is then a charged spherical shell with a thickness that remains much shorter than $b-a$ and with a mean radius $r_m$ which increases from $a$ to $b$ with time.

The potential and the electric field inside the capacitor have a spherical symmetry and they can be easily calculated imposing the proper boundary conditions and the continuity of the potential across the spherical shell constituting the photoelectron cloud. The electron spherical shell has a small but finite thickness. Electrons on the inner surface are subject to a force $F^-$ which is different from the force $F^+$ exerted on the electrons on the external surface. Then the work done by the electric forces during the flight is different for the two classes of electrons and this results in a difference in final kinetic energy given by
\begin{eqnarray}
\Delta E & = & \int_a^b (F^+-F^-) \, dr_m = \int_a^b \frac{-e Q}{4 \pi \varepsilon_0 r_m^2} \, dr_m = \nonumber \\
         & = & \frac{-e Q}{4 \pi \varepsilon_0}\frac{b-a}{ab}
\label{eq:LIK_ab}
\end{eqnarray}
where $-e$ is the electron charge and $Q$ is the charge of all the other photoelectrons. Usually the distance between the sample and the detector is much larger than the linear dimensions of the light spot and then assuming $b \gg a$
$$
\Delta E \approx \frac{-e Q}{4 \pi \varepsilon_0 a}.
$$
Nevertheless, in a typical PES experiment electrons are emitted only from one side of the sample. Then we want to relate the broadening to the number $N$ of electrons emitted only into half of the space. If $2N$ is the total number of electrons emitted from the sphere and we assume $2N \gg 1$, we have $Q=-(2N-1)e\approx -2Ne$ and the energy broadening results
\begin{equation}
\Delta E \approx \frac{e^2}{2 \pi \varepsilon_0}\frac{N}{a} = 2.88 \cdot 10^{-3} \mbox{eV} \cdot \mu\mbox{m}\frac{N}{a}
\label{eq:LIK_broadening}
\end{equation}
which is the prediction of the LIK model. The energy broadening $\Delta E$ is directly proportional to the number $N$ of photoelectrons and inversely proportional to the linear size $a$ of the spot on the sample surface. It does not depend upon the initial kinetic energy $E_0$ and upon the pulse duration $\tau_p$ when the latter is much smaller than the electrons' time of flight. Because even fast electrons with a kinetic energy of 10~keV take 1.7$\cdot$10$^{-8}$~s to cover a distance of 1~m, this last condition is easily verified with modern femtosecond-pulsed sources. Carrying out the calculations, it is found that the average final kinetic energy of the photoelectrons results $\langle E \rangle =E_0+e V_0$, the value expected from the conservation of energy.

\begin{table*}
\small
\centering
\begin{tabular}{l@{\hspace{0.1 cm}} l@{\hspace{0.1 cm}} r@{\hspace{0.2 cm}} r@{\hspace{0.3 cm}} d{3}@{\hspace{0.6 cm}} d{3}@{\hspace{0.6 cm}}}
\hline
\hline
Article                                    & PES structure          & \multicolumn{1}{c@{\hspace{0.1 cm}}}{$N$}          & \multicolumn{1}{c@{\hspace{0.0 cm}}}{$a$ ($\mu$m)}      &\multicolumn{1}{c@{\hspace{0.0 cm}}}{Exp. $\Delta E$ (eV)}    &  \multicolumn{1}{c}{LIK $\Delta E$ (eV)} \\
\hline
Zhou 2005 \cite{Zhou_05}           & Fermi edge             & 1875         & 180               & 0.015                  & 0.031 \\
Hellmann 2012 \cite{Hellmann_12a}  & Ta 4$f_{7/2}$          & 60000        & 250               & 0.35                   & 0.72  \\
Faure 2012 \cite{Faure_12}         & Fermi edge             & 2400         & 25                & 0.22                  & 0.28  \\
\hline
\hline
\end{tabular}
\caption{Comparison between the measured broadening of PES structures reported in some articles and the predictions provided by the LIK model. Zhou 2005 describes an experiment at a 3rd generation synchrotron facility on polycrystalline gold at 20~K with photon energy $h\nu$=35~eV, pulse duration of 60~ps and distance between two pulses of 2~ns. Hellmann 2012 refers to an experiment at a free-electron laser (FLASH) on 1T-TaS$_2$ sample at 10~K with $h\nu$=156~eV and macrobunches of 30 pulses separated by 4~$\mu$s, each pulse with a duration of $\sim$100~fs, 5 macrobunches per second. Faure 2012 describes an experiment on polycrystalline copper at 35~K using a Ti:Sapphire laser with $h\nu$=6.28~eV (fourth harmonic), pulse duration of 67.5~fs and a pulse separation of 4~$\mu$s.}
\label{tab:LIK_papers}
\end{table*}

Despite its exceptional simplifications, the LIK model succeeds in giving the correct order of magnitude for the  measured energy broadening in most experiments. Long et al. had already found a correspondence between their model and published data for tr-PES experiments \cite{Long_96}. In Table~\ref{tab:LIK_papers} we report the results of three recent articles which present a study of energy broadening as a function of the photoemitted electrons per pulse $N$. $N$ is extracted directly from sample current measurements or from the value of the photon flux  multiplied by the quantum efficiency $\eta$ of the material. In Table~\ref{tab:LIK_papers} we report the value of $N$ corresponding to the maximum sample current or photon flux, the linear dimension $a$ of the photon spot and the energy broadening $\Delta E$ due to space-charge effects, which is measured from the experimental spectra subtracting in quadrature the linewidth obtained at very low photon flux. The value of $\Delta E$ expected from LIK model is calculated through formula (\ref{eq:LIK_broadening}) and is reported in the last column. The LIK model tends to overestimate the experimental energy broadening but the order of magnitude is well reproduced. The three papers report a substantial linear dependence of $\Delta E$ from the sample current or the photon flux and Zhou et al.\cite{Zhou_05} also demonstrate an inverse proportionality with the spot size. The overestimation of the energy broadening can be ascribed to the fact that in the LIK model all the photoelectrons have the same energy whereas in a real experiment the electron cloud is dominated by secondary electrons which have a kinetic energy lower than the primary electrons that form the investigated PES structure. As pointed out by Zhou et al., electrons with similar energy tend to mutually interact more than electrons with different energy because they have higher probability to share a similar path during their flight. Also, in a real case secondary electrons tends to push forward faster electrons resulting in a shift to higher kinetic energy of the PES structures which increases linearly with $N$. This shift is not provided by the simple LIK model which treats mono-energetic electrons. We mention that a remarkable deviation from the LIK model is observed at very low electron kinetic energy ($\leq$1~eV) where the energy broadening $\Delta E$ is found proportional to the square root of the number $N$ of photoelectrons \cite{Passlack_06,Graf_10,Hellmann_09}.

We show for the first time that the LIK model can also offer an estimation of the time scale of the energy broadening effect. With reference to Fig.~\ref{fig:LIK_model}, when the electron spherical shell is at a distance $r_m$ from the center the energy broadening is (cfr. equation (\ref{eq:LIK_ab}))
\begin{equation}
\delta E (r_m)= \frac{N e^2}{2 \pi \varepsilon_0}\frac{r_m-a}{a r_m}
\label{eq:LIK_rm}
\end{equation}
where we used the relation $Q\approx-2Ne$ and we indicate with $\delta E$ the partial energy broadening during the flight. Now we want to express $r_m$ as a function of time. If the energy broadening remains much smaller than the initial kinetic energy $E_0$ and also $eV_0 \ll E_0$, the velocity $v_m$ of the central position of the electron cloud can be considered approximately constant and has value
$$
v_m \approx \sqrt{\frac{2 E_0}{m}}
$$
where $m$ is the electron mass. The electrons are all emitted at the instant $t=0$ and then $r_m=a+v_m t$. Equation (\ref{eq:LIK_rm}) can then be written as
\begin{equation}
\delta E (t)=\frac{e^2}{2 \pi \varepsilon_0}\frac{N}{a}\frac{v_m t}{1+v_m t}=\Delta E \frac{v_m t}{a+v_m t}
\label{eq:LIK_vm}
\end{equation}
where $\Delta E$ is the final energy broadening in the hypothesis $b \gg a$ indicated in equation (\ref{eq:LIK_broadening}). If we divide the numerator and denominator of the last term of equation (\ref{eq:LIK_vm}) by $a$ and define the characteristic time
\begin{equation}
\tau_c = a/v_m = a \sqrt{\frac{m}{2 E_0}},
\label{eq:LIK_chartime}
\end{equation}
we can write
\begin{equation}
\delta E(t) = \Delta E \frac{t/\tau_c}{1+t/\tau_c}.
\label{eq:LIK_timedep}
\end{equation}
For $t \gg \tau_c$ $ \delta E (t)$ tends to the value of formula (\ref{eq:LIK_broadening}). Equation (\ref{eq:LIK_chartime}) indicates that the time scale for energy broadening is directly proportional to the linear size of the beam spot and inversely proportional to the square root of the kinetic energy of the electrons. Within a time $\tau_C$ after the pulse the photoemitted electrons cover a distance equal to linear dimension $a$ of the photon spot.

\section{The simulation procedure} \label{sec:simulation}

The opensource Treecode software simulates the motion of particles interacting via Coulomb forces and was designed by its authors for the study of gravitational $N$-body problems. Its extension to the simulation of a cloud of interacting electrons required the modification of very few lines of the code. The code uses approximations to integrate the equations of motion. Here we give a very simplified and synthetic description of the calculations performed by the code in  the case of interacting electrons and we refer the reader to the article by Barnes and Hut \cite{Barnes_86} or to the software's webpage \cite{Treecode} for a comprehensive picture.

The position and velocity of any of the $N$ interacting particles at a given instant $t$ are assigned and are indicated as $\mathbf{r}_i$ and $\mathbf{v}_i$, $i=1,...,N$, respectively. A cube large enough to contain all the particles represents the space of the system. This cube is divided into eight cubic cells and  each cubic cell that contains more than one particle is divided into eight subcells. The process is repeated recursively until, at the lowest level, every cell contains just one or no particle. For each cell $p$ in the hierarchic structure the number of electrons $n_p$ contained in it and the position of their center of mass $\mathbf{R}_p$ are calculated. We want to describe the interaction of the $i$-th electron of the system with the other $N-1$ electrons. We consider an external cell $p$ of side $D_p$ . If the distance between the $i$-th electron and the center of mass of the cell $|\mathbf{r}_i-\mathbf{R}_p|$ is greater than $D_p$ all the electron contained in the cell are represented by an effective particle of charge $-n_p e$ and position $\mathbf{R}_p$ which gives a contribution to the potential acting on the $i$-th electron equal to
\begin{equation}
\phi_{ip}=k_C \frac{-n_p e}{|\mathbf{r}_i-\mathbf{R}_p|}
\label{eq:Treecode_phi}
\end{equation}
where $k_C=1/(4\pi\varepsilon_0)$ is the Coulomb's constant. If instead $|\mathbf{r}_i-\mathbf{R}_p|$ is lower than $D_p$ the eight subcells are considered. The acceleration of the $i$-th electron is given by
\begin{equation}
\mathbf{a}_i=-\sum_p \frac{e}{m} \phi_{ip} \frac{\mathbf{r}_i-\mathbf{R}_p}{|\mathbf{r}_i-\mathbf{R}_p|^2}
\label{eq:Treecode_a}
\end{equation}
where the sum is on the higher-level cells $p$ for which $|\mathbf{r}_i-\mathbf{R}_p|>D_p$. The cells in the summation must collectively contain all the  $N-1$ electrons other than the $i$-th with no superposition. If $\delta t$ is the leapfrog integration time, we can define an intermediate velocity
$$
\mathbf{v}_i \left( t+\frac{\delta t}{2} \right) =  \mathbf{v}_i(t)+\mathbf{a}_i \frac{\delta t}{2}
$$
and then the position and velocity at the new instant $t+\delta t$ are
$$
\mathbf{r}_i(t+\delta t) = \mathbf{r}_i(t)+ \mathbf{v}_i \left( t+\frac{\delta t}{2} \right) \delta t
$$
$$
\mathbf{v}_i(t+\delta t)=\mathbf{v}_i \left( t+\frac{\delta t}{2} \right)+\mathbf{a}_i \frac{\delta t}{2}.
$$
It is possible to demonstrate that the approximations of the Treecode software reduce the number of forces to calculate at every step $\delta t$ from $\frac{1}{2}N(N-1)$, needed in conventional $N$-body treatment, to a value that grows only as $N \log(N)$. This property decreases enormously the computation time when the number of particles goes up. The Treecode software does not consider relativistic effects. The Lorentz term $\gamma=1/\sqrt{1-\beta^2}$ for electrons with a kinetic energy of 10~keV (the maximum value we explored) is 1.02 and we can regard the non-relativistic approximation as satisfactory.

\begin{figure}
  \centering
  \includegraphics[width= 7.5 cm]{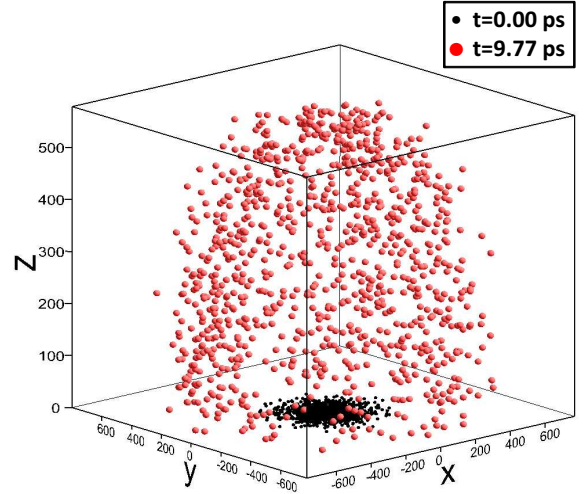}\\
  \caption{Positions of the photoelectrons obtained for a simulation with $N$=1000 electrons, a spot radius $a$=100~$\mu$m and an initial kinetic energy $E_0$=10~keV. Black dots represent electrons' positions at $t$=0 and red dots are the positions at $t$=9.77$\cdot 10^{-3}$~ns. The leapfrog integration time is $\delta t$=6.1$\cdot 10^{-5}$~ns. The lengths on the axes are expressed in $\mu$m. The dashed circle is a guide to the eye.}
  \label{fig:explosion}
\end{figure}

In our simulations at the instant $t=0$ the $N$ electrons are located on the plane $z=0$ which represents the sample surface. In order to better reproduce the experimental conditions, the electrons have a probabilistic bi-dimensional Gaussian distribution around the origin with a standard deviation $a$, which describes the linear dimensions of the photon beam spot and can be assumed as its ``radius''. The direction of their velocities is randomly distributed in the $2 \pi$ solid angle subtending the upper hemisphere ($v_z>0$) while the modulus obviously depends on the initial kinetic energy $E_0$. The numerical values of the initial positions and velocities and of the various simulation parameters are inserted scaling the fundamental units and expressing masses in nanograms (ng), lengths in micrometers ($\mu$m) and times in nanoseconds (ns) while electric charges are considered in coulombs. With this choice in equation (\ref{eq:Treecode_phi}) the quantity $k_C e$ is equal to $1.44\cdot 10^3 \mbox{ ng}\cdot\mu\mbox{m}^3\cdot\mbox{ns}^{-2}\cdot\mbox{C}^{-1}$ and in equation (\ref{eq:Treecode_a}) the quantity $e/m$ results $1.76\cdot 10^{-1} \mbox{ C}\cdot\mbox{ng}^{-1}$. In this way the program mostly works with numbers of order of magnitude close to unity. In the output file the Treecode software reports the position, velocity, acceleration and potential of each electron at various steps of the simulation. In Fig.~\ref{fig:explosion} the simulated position of 1000 electrons emerging from a spot of radius $a$=100~$\mu$m with an initial kinetic energy of 10,000~eV are reported at $t=0$ and at $t$=9.77$\cdot 10^{-3}$~ns after the pulse. From the output file we easily calculate the kinetic energy as a function of time $E_i(t)$ for each electron  and its shift with respect to the initial value at $t=0$ $\epsilon_i(t)=E_i(t)-E_i(0)$.

The leapfrog integration time $\delta t$ is a simulation parameter which must be treated with care. If it is taken too long, the obtained results may strongly differ from the real integrated solutions but a too short $\delta t$ remarkably increases the computation time. We have verified the reliability of our simulations employing decreasing values of $\delta t$. For sufficiently short $\delta t$ a further decrease has negligible effects on the final results. The conservation of total energy is a further test to check that a reliable integration time has been chosen. Varying the initial kinetic energy $E_0$ and the radius $a$ the leapfrog integration time is scaled according to the characteristic time $\tau_C$ as expressed in equation (\ref{eq:LIK_chartime}). As shown later, the energy broadening mainly occurs in an initial range of time equivalent to few times $\tau_C$. In order to save time, in some cases we used a smaller $\delta t$ in the first part of the simulation and we completed the procedure with a larger integration time.

Different extractions of random positions and velocities may lead to slightly different results, in particular if the number $N$ of electrons is small. In many cases we employed more initial configurations obtained by different random extractions and we averaged the results of the corresponding simulation procedures.

\section{Results of the simulation} \label{sec:results}

\subsection{Simulations with mono-energetic electrons} \label{sec:mono}

\begin{figure}
  \centering
  \includegraphics[width= 7.0 cm]{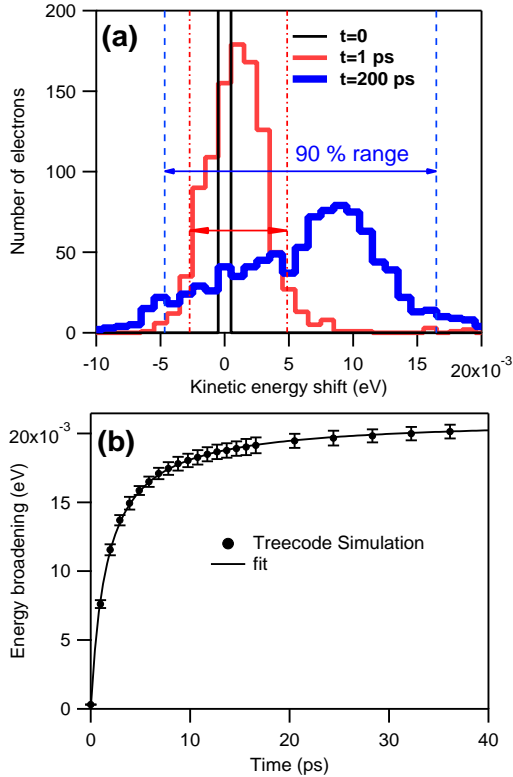}\\
  \caption{Simulation with $N$=1000 photoelectrons, spot radius $a$=100~$\mu$m and initial kinetic energy $E_0$=10,000~eV. {\em (a)}: Distribution of the kinetic-energy shift for the electrons at $t$=0 (thin black line), $t$=1~ps (medium red line) and $t$=200~ps (thick blue line). The top of the histogram bar at $t$=0 corresponds to 1000 and is out of the scale of the graph. The range between the 5th and the 95th percentile of the kinetic energy distribution (which we define as our energy broadening) is indicated by two vertical lines and results 7.6~meV at $t$=1~ps (red dash-dot lines) and 21.1~meV at $t$=200~ps (blue dashed lines). {\em (b)}: Variation of the energy broadening with time (solid circles) and its fit (solid line) using the function in equation (\ref{eq:LIK_timedep}).}
  \label{fig:broadening_time}
\end{figure}

The histogram in the panel (a) of Fig.~\ref{fig:broadening_time} reports the distribution of the kinetic energy shift $\epsilon_i$ for $N$=1000 electrons emerging from a spot of radius $a$=100~$\mu$m with the same initial kinetic energy $E_0$=10,000~eV at the three different instants $t$=0, 1~ps and 200~ps.  As expected, a broadening of the energy distribution occurs due to the Coulomb interactions between electrons. A shift in the average energy position is evident in particular at $t$=200~ps and is due to the conversion of the initial potential electrostatic energy of the electron cloud into kinetic energy when the particles drift apart (cfr. Fig.~\ref{fig:explosion}). Because we are treating the case in which all the electrons have at $t$=0 the same energy, this shift would not be present if we supposed that the sample plane is an equipotential surface and took into account the mirror charges of the photoelectrons. After just 1~ps the energy broadening is about one third of that encountered at $t$=200~ps. It is interesting to note that the distribution of energies after relatively long times does not resemble a Gaussian shape at all and presents an evident negative skewness. Moreover, a very small number of electrons is subject to energy shifts significantly greater than $\Delta E$ (we found $|\epsilon_i|$=0.12~eV for one particle) and are not reported in the range of the graph. These outliers can be attributed to short-range scattering between electrons \cite{Hellmann_09} which is not properly considered in the calculation. The number of these outliers tends to decrease shortening the leapfrog integration time $\delta t$. In order to avoid the contribution of the tails in the energy distribution and of the short-range scattered electrons, we define the energy broadening $\Delta E$ as the difference between the 5th and the 95th percentile in the energy distribution considering a central range containing 90\% of the total photoelectrons. The positions of the two extremes of this range are indicated in Fig.~\ref{fig:broadening_time}(a) for $t$=1~ps and $t$=200~ps.

In the panel~(b) of Fig.~\ref{fig:broadening_time} the partial energy broadening $\delta E$ is reported as a function of time between $t$=0 and $t$=40~ps. We considered 10 different initial configurations of electron positions and velocities. In the graph we report the average of the calculated values of $\delta E(t)$ for the 10 initial configurations and the error bars represent their standard deviation. We fitted the simulated $\delta E(t)$ with the expression in equation (\ref{eq:LIK_timedep}) considering $\Delta E$ and $\tau_C$ as fitting parameters. The expression obtained from LIK model perfectly reproduces the simulation with best fit parameters $\Delta E$=21.06$\pm$0.07~meV and $\tau_C$=1.64$\pm$0.05~ps. The value expected by the LIK model for $\Delta E$ through equation (\ref{eq:LIK_broadening}) results 28.8~meV and is quite similar to the value found in the fitting. LIK model through equation (\ref{eq:LIK_chartime}) provides $\tau_C$=1.69~ps which corresponds to the fitted value within the error.

\begin{figure}
  \centering
  \includegraphics[width= 8 cm]{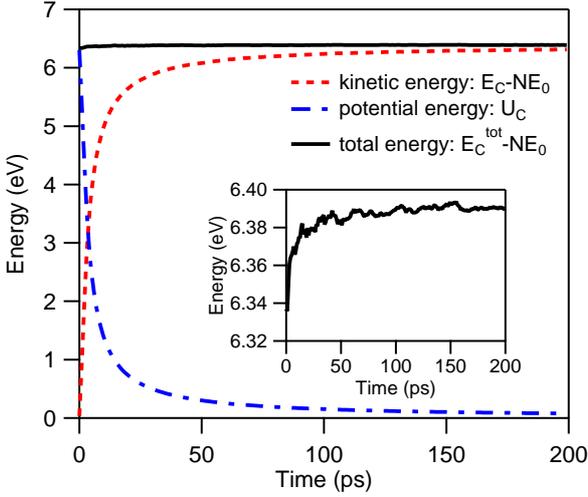}\\
  \caption{Simulation with $N$=1000 photoelectrons, spot radius $a$=100~$\mu$m and initial kinetic energy $E_0$=10,000~eV. $E_C-NE_0$ (red dashed line), $U_C$ (blue dash-dot line) and $E_C^{tot}-NE_0$ (black solid line) are plotted as a function of time. In the inset we report a magnification for the vertical axis which better illustrates the variation of total energy with time.}
  \label{fig:energy}
\end{figure}

In order to test the reliability of our simulation we can check the conservation of energy for the system of interacting particles during the flight. The total kinetic energy of the overall electron cloud is given by
$$
E_C(t)=\sum_{i=1}^{N}E_i(t)=\sum_{i=1}^{N}\epsilon_i(t) + N E_0
$$
where $N E_0$ is the kinetic energy of the electron cloud at $t$=0. The potential acting on the $i$-th electron is calculated as
$$
\phi_i(t)=\sum_p \phi_{ip}(t)
$$
where $\phi_{ip}$ is the contribution of the $p$-th cell in the hierarchic structure defined in equation (\ref{eq:Treecode_phi}) and the summation is extended to the higher levels cells $p$ which obey the condition $|\mathbf{r}_i-\mathbf{R}_p|>D_p$, as in equation (\ref{eq:Treecode_a}). The potential energy of the electron cloud is then given by
$$
U_C(t)=\frac{1}{2}(-e)\sum_{i=1}^N \phi_i(t).
$$
The total energy of the electron cloud $E_C^{tot}(t)=E_C(t)+U_C(t)$ is expected to be constant with time. In Fig.~\ref{fig:energy} we report the simulated quantities $E_C-NE_0$, $U_C$ and $E_C^{tot}-NE_0$ as a function of time. In the graph we have subtracted the initial kinetic energy $N E_0$=10$^7$~eV to the kinetic and to the total energies for convenience. The electrostatic potential energy decreases with time due to the spread of photoelectron cloud and is converted into an increase of the kinetic energy. The variation of the total energy is about 55~meV (see inset), which corresponds to less than 1\% of the quantity $E_C^{tot}-NE_0$. If we refer to $E_C^{tot}$ with no subtraction of $NE_0$, the total energy is conserved in the simulation within a tolerance of less than one part on 100~millions.

\begin{figure}
  \centering
 \includegraphics[width=8.8 cm]{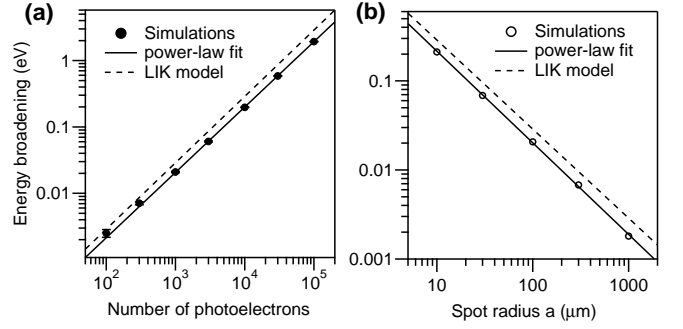}\\
  \caption{Simulations with mono-energetic photoelectrons with $E_0$=10,000~eV. {\em (a)}: Energy broadening (solid dots) as a function of the number $N$ of mono-energetic photoelectrons, power-law fit of the type $\Delta E(N)=C N^\alpha\ /a$ (solid line) and energy broadening expected from LIK model (dashed line). The spot radius $a$ is fixed at 100~$\mu$m. {\em (b)}: Energy broadening values for $N$=1000 photoelectrons (empty dots) as a function of spot radius $a$, power-law fit of the type $\Delta E(a)=C N/a^{\beta}$ (solid line) and the prediction of the LIK model (dashed line).
  In both panels, the energy broadening values are calculated as the average of 10 simulations with different initial configurations (5 configurations for $N$=5$\cdot$10$^4$ and $N$=2$\cdot$10$^5$) and the error bars, when greater of the mark dimension, represent the standard deviation.}
  \label{fig:LIK_fit}
\end{figure}

We performed a set of simulations varying the number $N$ of photoemitted electrons in a range from 100 to 100,000 and keeping constant the size $a$=100~$\mu$m of the spot and the initial kinetic energy $E_0$=10,000~eV. In the panel (a) of Fig.~\ref{fig:LIK_fit} the solid dots represent the energy broadening $\Delta E$ as a function of $N$, calculated at a time $t$= 250~ps, much greater than the characteristic time $\tau_C\simeq$1.64~ps. The reported values are obtained averaging the results of several simulations with different initial configurations of electrons' position and velocity. The points align along a straight line in the log-log scale. We fit the simulated values of energy broadening with a power law of the type $\Delta E(N)=C N^\alpha\ /a$. The solid line in Fig.~\ref{fig:LIK_fit}~(a) is the best fitting power curve and corresponds to the best parameters $C=(2.27\pm0.07)\cdot 10^{-3}\mbox{ eV}\cdot\mu\mbox{m}$ and $\alpha=0.985\pm0.003$. The LIK model provides $C_{LIK}=2.88 \cdot 10^{-3}\mbox{ eV}\cdot\mu\mbox{m}$ and $\alpha_{LIK}$=1. The direct proportionality of $\Delta E$ to $N$ provided by the LIK model in equation (\ref{eq:LIK_broadening}) is well reproduced by the simulations. The proportionality constant is slightly lower than expected and the broadening provided by the LIK model is reported by the dashed line.

The same set of simulations has been carried out with photoelectrons with an initial kinetic energy of $E_0$=100~eV and $a$=100~$\mu$m. Despite the different kinetic energy and the different characteristic time, which results in this case $\tau_C$=16.4~ps,  the simulated values of energy broadening for the same $N$ are almost identical to the higher energy case and the linear dependence of $\Delta E$ with $N$ is confirmed. Moreover, this result is in agreement  with the predictions of the LIK model according to which $\Delta E$ is determined by $N$ and $a$ but is independent of the initial kinetic energy $E_0$.

We also checked the dependence of $\Delta E$ from the spot size $a$ carrying out simulations with $N$=1000, $E_0$=10000~eV and $a$ varying in the range $a$=10--1000~$\mu$m. For the dependence of the characteristic time from $a$, different leapfrog integration times were used and the energy broadening values are always calculated for $t \gg \tau_C$. Simulated values of $\Delta E$ as a function of $a$ are reported in Fig.~\ref{fig:LIK_fit}~(b) and are fitted with a power law of type $\Delta E(a)=C N/a^{\beta}$. The best fitting parameters are in this case $C=(2.30\pm 0.01) \cdot 10^{-3}\mbox{ eV}\cdot\mu\mbox{m}$ and $\beta=1.029\pm 0.009$. The inverse proportionality of $\Delta E$ with respect to $a$ is well verified and the prefactor $C$ is compatible within the error with that obtained in the previous set of simulations where the initial kinetic energy was varied.

\subsection{Simulations of 2p peaks from metallic copper} \label{sec:Cu}

We consider now the space-charge effects in a simple realistic case. We simulate a pulsed high-energy photoemission experiment from pure polycrystalline Cu and we want to  investigate the broadening effect on the 2p peaks. Metallic Cu is one of the few materials for which the quantum efficiency $\eta$, i.e. the ratio between the number of the overall photoemitted electrons and the number of impinging photons, is known in literature also at relatively high photon energies \cite{Day_81,Henneken_00a}. We perform our simulations for the two different  values of photon energy $h\nu$=8000~eV and $h\nu$=9000~eV, the former below and the latter just above the Cu K absorption edge. Cu quantum efficiency results $\approx$0.004 at 8000~eV and $\approx$0.015 at 9000~eV. Secondary electrons constitutes the larger part of the total photoemitted particle (80--90\% for typical metals at keV energies \cite{Henke_81}), the rest of the spectrum being composed by core-level primary electrons, Auger electrons, primary electrons emitted from valence bands and high-energy inelastic background.

The number of core-level photoelectrons per incident photon is a key parameter in the feasibility study of PES experiments. Fadley gave an expression for the expected number of detected photoelectrons from a subshell $c$ of a given element emitted from a semi-infinite homogeneous specimen (equation~115 in reference~\cite{Fadley_78})
\begin{equation}
n_c=I_0 \Omega_0(E_c) A_0(E_c) D_0(E_c) \rho \frac{d\sigma_c (h\nu)}{d \Omega} \lambda_e(E_c),
\label{eq:fadley}
\end{equation}
where $I_0$ is the photon flux, $E_c$ is the kinetic energy of photoelectrons from core-level $c$, $\Omega_0$ is the effective acceptance solid angle of the detector, $A_0$ is the effective area of the sample, $D_0$ is the detector efficiency, $\rho$ is the density of atoms of the analyzed element, $d \sigma_c / d \Omega$ is the differential cross section for photoemission from the core-level $c$ and $\lambda_e$ is the inelastic mean free path for electrons in the considered material. As mentioned above, $\lambda_e$ has a mild dependence on $E_c$, but also $\Omega_0$, $A_0$ and $D_0$ may depend on the photoelectron kinetic energy. We want to estimate the number of 2p$_{1/2}$ and 2p$_{3/2}$ photoelectrons generated by X-ray radiation normally incident on a sample of pure polycrystalline copper and emitted from the surface in the upper half-space. We can use Fadley's expression (\ref{eq:fadley}) with $\Omega_0$=2$\pi$, $D_0$=1 and assume $A_0$ as the the area of the photon spot on sample surface. Because the sample is polycrystalline we can ignore the dependence of photoemission probability on emission angle and write $d \sigma_c / d \Omega$=$\sigma_c/4\pi$, where $\sigma_c$ is the total photoabsorption cross section. If $N_{ph}=I_0 A_0$ is the number of photons per pulse impinging on the sample we can finally write
\begin{equation}
n_c \approx \frac{1}{2}\sigma_c(h\nu)\rho\lambda_e(E_c)N_{ph}
\label{eq:photoemission}
\end{equation}
The kinetic energy of the electrons emitted from the core level $c$ is given by the relation $E_c=h\nu-E_{Bc}-\phi_W$, where $E_{Bc}$ is the binding energy of level $c$ and $\phi_W$ is the work function of copper. This last depends strongly on the crystallographic termination of the surface but we can assume $\phi_W$=4.8~eV as an intermediate value among various terminal planes \cite{CRC_manual_05}. We use formula (\ref{eq:photoemission}) to calculate the ratio between the expected number of photoelectrons $n_c$ and the number of the incident photons $N_{ph}$ for the Cu 2p$_{1/2}$ and 2p$_{3/2}$ peaks  in simulated experiments with $h\nu$=8~keV and 9~keV. The  parameters used in formula (\ref{eq:photoemission}) can be found in literature \cite{XDB_09,Scofield_report_73, IMFP, Powell_99} and are listed in Table~\ref{tab:Cu_2p} with the calculated $n_c/N_{ph}$ ratios.

\begin{table}
  \centering
  \small
  \setlength{\tabcolsep}{5pt}
\begin{tabular}{l *{4}{p{1.42 cm}}}
\hline
\hline
            &   \multicolumn{2}{c}{$h\nu$=8000 eV} &   \multicolumn{2}{c}{$h\nu$=9000 eV} \\
            & Cu 2p$_{1/2}$     & Cu 2p$_{3/2}$     &  Cu 2p$_{1/2}$        & Cu 2p$_{3/2}$ \\
\hline
$E_{Bc}$ (eV)  &    952.3   & 932.7                 &   952.3               &   932.7   \\
$E_c $ (eV)    & 7042.9       & 7062.5            &   8042.9              & 8062.5    \\
$\sigma_c$ (barns)& 7.70$\cdot$10$^2$ & 1.415$\cdot$10$^3$   &   5.20$\cdot$10$^2$    &   9.51$\cdot$10$^2$\\
$\lambda_e$ (\AA) & 78.07               & 78.25                 & 87.00                 & 87.17\\
$n_c/N_{ph}$    & 2.55$\cdot$10$^{-5}$ & 4.70 $\cdot$10$^{-5}$    &   1.92$\cdot$10$^{-5}$ &   3.52$\cdot$10$^{-5}$\\
\hline
\hline
\end{tabular}
  \caption{The ratios $n_c/N_{ph}$ for Cu 2p$_{1/2}$ and 2p$_{3/2}$ peaks for a pure Cu sample using $h\nu$=8000~eV and $h\nu$=9000~eV radiation and the parameters to calculate them through the formula~(\ref{eq:photoemission}). Binding energy values $E_{Bc}$ are obtained from the X-ray Data Booklet \cite{XDB_09}. Photoemission cross sections $\sigma_c$ are obtained from Scofield \cite{Scofield_report_73}. $\lambda{}_e $ values are calculated through the software NIST Electron Inelastic-Mean-Free-Path Database \cite{IMFP} using the parameters by Powell and Jablonski \cite{Powell_99}.}
  \label{tab:Cu_2p}
\end{table}

The space-charge effects on 2p electrons is mainly due to secondary electrons whose number $n_{\mbox{sec}}$ is approxi\-mate\-ly given by the quantum efficiency $\eta$ multiplied by the number of imping photons per pulse $N_{ph}$. For simplicity, we consider the secondary electrons to be mono-energetic with an initial kinetic energy equal to the peak of the secondary electron distribution. For the simulations, we are interested to the kinetic energy referred to vacuum level, that is the kinetic energy of the photoelectrons after they have overcome the potential barrier represented by the work function $\phi_W$ and left the surface. In metals, the peak of the distribution of the kinetic energy of secondary electrons is about one third of $\phi_W$ \cite{Chung_MS_74}, and so we can assume 1.6~eV as the initial energy of the secondary electrons in the simulation. Also the electrons of the Cu 2p$_{1/2}$ and 2p$_{3/2}$ peaks are considered as emitted with the same energies $E_c$ listed in Table~\ref{tab:Cu_2p} ignoring the natural linewidth due to finite hole lifetime. The final broadening at the end of the simulation is so due purely to space-charge effects. We ignore the contribution of other core-level and Auger peaks because they are distant in energy respect to the 2p lines and are much less numerous than secondary electrons. The numbers $n_{2p1/2}$ and $n_{2p3/2}$ of 2p$_{1/2}$ and 2p$_{3/2}$ electrons, respectively, are obviously calculated through equation (\ref{eq:photoemission}). Let us consider for example the case of a number of $N_{ph}=5\cdot10^6$ photons per pulse, a value easily accessible in the X-ray FEL beamlines under construction \cite{Schneidmiller_techrep_11}. For $h\nu$=8000~eV our electron cloud in the simulation will be composed by $n_{\mbox{sec}}$=20,000 secondary electrons with an initial kinetic energy $E_c$=1.6~eV, $n_{2p1/2}$=128 electrons with $E_c$=7042.9~eV and $n_{2p3/2}$=235 electrons with $E_c$=7062.5~eV. For $h\nu$=9000~eV we have $n_{\mbox{sec}}$=75,000 secondary electrons with an initial kinetic energy of 1.6~eV, $n_{2p1/2}$=96 electrons with $E_c$=8042.9~eV and $n_{2p3/2}$=176 electrons with $E_c$=8062.5~eV. We note that increasing $h\nu$ from 8000 to 9000~eV the number of 2p photoelectrons slightly varies but secondary electrons remarkably grow in number for the new generation channels due to 1s photoelectrons and K-shell Auger electrons, obviously absent below the Cu K absorption edge.

\begin{figure}
  \centering
  \includegraphics[width=8.8 cm]{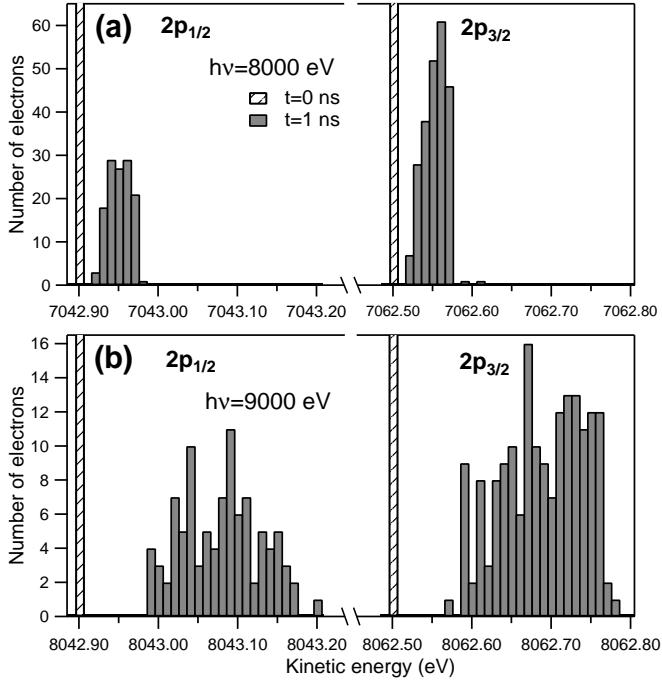}\\
  \caption{Energy distribution of Cu 2p photoelectrons from pure metallic copper at $t$=0 (patterned bar) and $t$=1~ns (full bars) in the simulation with $h\nu$=8000~eV (panel~(a)) and in the simulation with $h\nu$=9000~eV (panel~(b)). For both the simulations $N_{ph}$=5$\cdot$10$^6$ and $a$=500~$\mu$m. The top of the histogram bars for $t$=0 is out of the scale of the graph.}
  \label{fig:Cu_2p_hist}
\end{figure}

In Fig.~\ref{fig:Cu_2p_hist} we report the energy distribution at $t$=0 and at $t$=1~ns for 2p$_{1/2}$ and 2p$_{3/2}$ peaks obtained in two simulations with $h\nu$=8000~eV (panel~(a)) and $h\nu$=9000~eV (panel~(b)), with $N_{ph}$=5$\cdot$10$^6$ and $a$=500~$\mu$m for both the simulations. It is immediately evident the much greater energy broadening and peak dispersion for $h\nu$=9~keV (we find $\Delta E$=164~meV and 166~meV for 2p$_{1/2}$ and 2p$_{3/2}$, respectively) if compared to the case $h\nu$=8~keV ($\Delta E$=41~meV and 44~meV). This can be easily ascribed to the much greater number of secondary electrons present at higher photon energy and emphasizes the influence that the choice of $h\nu$ in an experiment may have on the secondary electrons' production and on the energy broadening induced by the space charge. The LIK model predicts an energy broadening $\Delta E$=434~meV for the $h\nu$=9~keV experiment and $\Delta E$=117~meV for the $h\nu$=8~keV experiment, in both cases almost three times the values found in the simulations. In the calculations with mono-energetic electrons the discrepancy between many-body simulations and the LIK model was much lower. This is attributable to the very different velocities of core-level and secondary electrons so that the two families of electrons tend to drift apart. Core-level electrons interact with the secondary electrons for a shorter time with respect to the case of a single-energy photoelectron cloud, resulting in a smaller broadening. We also note that we have a considerable shift of the final average kinetic energy with respect to the initial values at $t$=0, resulting for 2p$_{3/2}$ electrons in 185~meV for the $h\nu$=9~keV experiment and in 54~meV for the $h\nu$=8~keV experiment. Both these values are slightly larger than the respective energy broadening $\Delta E$ whereas in the previous case of mono-energetic electrons the energy shift was about one third of the broadening. Also in this case, the difference can be ascribed to the presence of the numerous secondary electrons which push forward the faster Cu 2p electrons and it is consistent with the results of lower-energy experiments where the energy shift has usually values similar to the energy broadening \cite{Zhou_05}.

\begin{figure}
  \centering
  \includegraphics[width=7.0 cm]{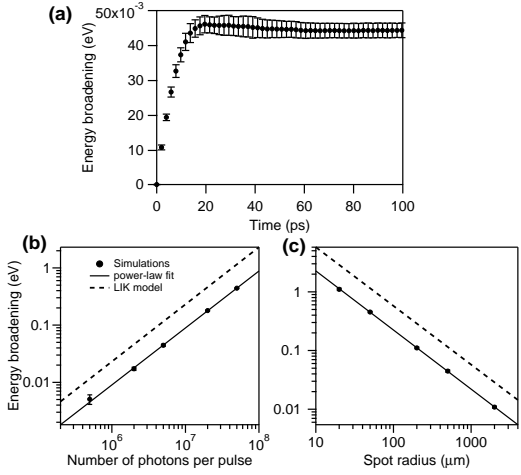}\\
  \caption{Many-body simulation of photoemission from pure Cu with $h\nu$=8000~eV. {\em (a)}: Energy broadening of the 2p$_{3/2}$ peak as a function of time for $N_{ph}$=5$\cdot$10$^6$ and $a$=500~$\mu$m. The reported curve is the average of the results from 10 initial configurations and the error bars represent the standard deviation. {\em (b)-(c)}: Energy broadening of the 2p$_{3/2}$ peak (solid dots) as a function of the number of photons $N_{ph}$ for $a$=500~$\mu$m (panel (b)) and as a function of the spot radius $a$ for $N_{ph}$=5$\cdot$10$^6$ (panel (c)). The reported values for different $N_{ph}$ and $a$ are calculated at a time $t\gg\tau_C$ and are the averages obtained from 10 initial configurations (5 configurations for $N_{ph}$=2$\cdot$10$^7$ and 5$\cdot$10$^7$). The power-law fit (solid lines) and the values expected from LIK model (dashed lines) are also reported.}
  \label{fig:Cu_2p_LIK}
\end{figure}

In Fig.~\ref{fig:Cu_2p_LIK}~(a) we report the evolution with time of the energy broadening of the Cu 2p$_{3/2}$ peak for $h\nu$=8000~eV, $N_{ph}=5\cdot10^6$ and $a$=500~$\mu$m. The broadening of the electrons' energy distribution takes place in the first 20~ps and for the remaining part of the simulation $\delta E$ stays approximately constant within the error. The initial kinetic energy of Cu $2p_{3/2}$ electrons being $E_0$=7062.5~eV, from equation (\ref{eq:LIK_chartime}) we obtain a characteristic time $\tau_c$=10~ps, which is comparable with the dynamics of the energy broadening mechanism obtained in the simulations. As stressed at the end of Section~\ref{sec:LIK}, at $t=\tau_C$ the 2p photoelectrons have travelled from the sample surface at a distance of the order of the spot dimension, in the present case $a$=500~$\mu$m. Because the size of the beam spot is usually much smaller than the distance between the sample and the detector, the energy broadening occurs in the very initial part of the photoelectrons' flight.

In Fig.~\ref{fig:Cu_2p_LIK}~(b) the simulated energy broadening for Cu 2p$_{3/2}$ electrons at time $t$=1~ns is reported as a function of $N_{ph}$ for fixed $a$=500~$\mu$m. $N_{ph}$ varies between $5 \cdot 10^5$ and $5 \cdot 10^7$, this corresponding to a number of photoelectrons between approximately $2 \cdot 10^3$ and $2 \cdot 10^5$. In Fig.~\ref{fig:Cu_2p_LIK}~(c) we report the energy broadening of Cu 2p$_{3/2}$ electrons as a function of spot radius $a$ (20--2000~$\mu$m) for a fixed number of impinging photons per pulse $N_{ph}=5\cdot 10^6$. These energy broadening values are calculated at a time $t\gg\tau_C$, where $\tau_C$ id obtained from equation~(\ref{eq:LIK_chartime}). The dependence of $\Delta E$ on $N_{ph}$ and $a$ are evidently linear in the log-log scale. We tried to fit the data points of energy broadening as a function of $N_{ph}$ in Fig.~\ref{fig:Cu_2p_LIK}~(b) with a power law of the type $\Delta E(N_{ph})=D (N_{ph})^\alpha /a$ with $D$ and $\alpha$ as fitting parameters. The best fitting power law is the solid line in the panel and is characterized by $D=(4.8 \pm 1.1) \cdot 10^{-6}$~eV$\cdot \mu$m and $\alpha$=0.996$\pm$0.014. From LIK model we would expect $D_{LIK}=C_{LIK} \eta =1.15\cdot 10^{-5}$~eV$\cdot \mu$m and $\alpha_{LIK}$=1. We also fitted the data point in Fig.~\ref{fig:Cu_2p_LIK}~(c) with a power law $\Delta E (a)=D N_{ph}/a^\beta$ obtaining as best fit parameters $D=(4.61\pm0.27)\cdot 10^{-6}$ and $\beta$=1.007$\pm$0.011. Very similar results are obtained for Cu 2p$_{1/2}$ electrons. We obtain that, also in the simulation of a realistic experiment, the energy broadening is directly proportional to the number of impinging photons per pulse (and then to the total number of photoelectrons per pulse) and inversely proportional to the linear dimension of the spot on the sample surface. The linear dependence of the energy broadening  on the photon flux and on the number of photoelectrons per pulse has been evidenced in most experiments that investigate PES structures with a kinetic energy greater than few eV \cite{Zhou_05, Hellmann_12a, Pietzsch_08,Buckanie_09,Faure_12}. Very recently Oloff~et~al. carried out at the SACLA FEL at Spring-8 an investigation on relatively high energy (2--3~keV) core-level photoelectrons from VO$_2$ and SrTiO$_3$ and found an unexpected non-linear dependence of the induced energy broadening on the photon flux \cite{Oloff_14}. They attributed this discrepancy to a non-isotropic electron emission and anomalies in photoelectron energy distribution and stressed that their measurements are affected by large experimental uncertainties, especially in the determination of the photon flux. Moreover, the use of insulating samples may have caused charging effect on their surface.

\begin{figure}
  \centering
  \includegraphics[width=7.2 cm]{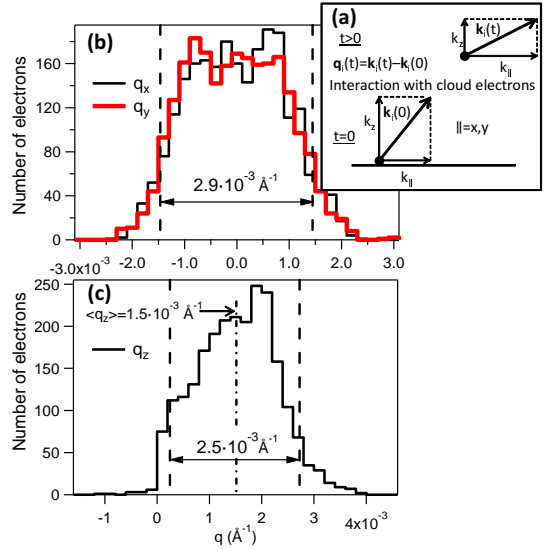}\\
  \caption{{\em (a)}: The momentum of a photoelectron can be modified by the interaction with other electrons and at long time $t$ after the emission from the surface its momentum $\mathbf{k}_i(t)$ is different in modulus and direction with respect to the initial value at $t$=0 $\mathbf{k}_i(0)$. $\mathbf{q}_i(t)=\mathbf{k}_i(t)-\mathbf{k}_i(0)$ is the variation of momentum for a single electron. {\em (b)}: Simulated distribution of the components of momentum variation parallel to sample surface $q_x$ (thin black solid line) and $q_y$ (thick red solid line) for Cu 2p$_{3/2}$ electrons in a photoemission experiment on metallic Cu with $N_{ph}$=5$\cdot$10$^7$ and $a$=500~$\mu$m at $t$=1~ns after the pulse. The two dashed vertical lines indicate the 90\% central range of the two distributions (very similar for the two components) which results 2.9$\cdot 10^{-3}$~\AA$^{-1}$ wide. {\em (c)}: Simulated distribution of the component of momentum variation perpendicular to sample surface $q_z$ (black solid line). The dash-dot vertical line indicates the average value of $q_z$ which is 1.5$\cdot 10^{-3}$~\AA$^{-1}$. The two dashed vertical lines indicate the 90\% central range of the distribution which results 2.5$\cdot 10^{-3}$~\AA$^{-1}$ wide.}
  \label{fig:Cu_2p_momentum}
\end{figure}

In order to perform ARPES experiments, it is also important to analyze how the mutual interactions among electrons can affect the direction of their velocities and not only their kinetic energy. If $\mathbf{v}_i$ is the velocity of the $i\mbox{-th}$ electron, its wavevector is given by $\mathbf{k}_i=m \mathbf{v}_i / \hbar$, which is also called electron's momentum in ARPES terminology. We then define the momentum variation of the $i$-th electron at the instant $t$ the vector
$$
\mathbf{q}_i(t)=\mathbf{k}_i(t)- \mathbf{k}_i(0)
$$
where $\mathbf{k}_i(t)$ and $\mathbf{k}_i(0)$ are, respectively, the momentum of the $i$-th electron at time $t$ and immediately after the photoemission. In absence of space-charge effects, the electrons would fly on a straight line and $\mathbf{k}_i$ would be constant with time. The space charge modifies $\mathbf{k}_i$ during the flight and introduces a non-null $\mathbf{q}_i$ (see Fig.~\ref{fig:Cu_2p_momentum}~(a)). We note that the sum of the momentum variation $\mathbf{q}_i(t)$ over all the electrons must have all its three components null for the conservation of total momentum.

Let us consider one of the ``worst'' cases examined above in terms of energy broadening. For $N_{ph}=5 \cdot 10^7$ and $a$=500~$\mu$m the expected energy broadening is 450~meV, a value which would make difficult a fine analysis of a photoemission peak structure. In Fig.~\ref{fig:Cu_2p_momentum}~(b) we report the distribution of the $x$ and $y$ components (parallel to sample surface) of the momentum variation $\mathbf{q}$ for the Cu 2p$_{3/2}$ electrons. The two distributions are very similar and symmetric with respect to the zero value and present a broadening (calculated as the width of the 90\% central range) of 2.9$\cdot10^{-3}$~\AA$^{-1}$. For the analysis of 3D structures also the component of electron momentum perpendicular to sample surface is important. The distribution of the $z$ component of $\mathbf{q}$ is shown in Fig.~\ref{fig:Cu_2p_momentum}~(c). This distribution is clearly shifted towards positive values of $q_z$ with an average value of 1.5$\cdot$10$^{-3}$~\AA$^{-1}$ and a 90\% central range 2.5$\cdot10^{-3}$~\AA$^{-1}$ wide. The shift of the distribution of $q_z$ towards positive values is attributable to the push effect from secondary electrons. We note that these values of momentum shift and broadening are much smaller than the modulus of the electron momentum for Cu 2p$_{3/2}$ electrons, which can be calculated as $|\mathbf{k}| \approx \sqrt{2 E_0 m}/\hbar$=43~\AA$^{-1}$. Most importantly, the broadening in the distribution of $q_x$, $q_y$ and $q_z$ results much smaller than the typical values for the linear dimension of the Brillouin zone (few~\AA$^{-1}$) and is not expected to affect the usually requested momentum resolution of ARPES experiments. In a case of reduced spot dimension, $a$=20~$\mu$m and $N_{ph}=5\cdot 10^6$, in face of an energy broadening of 1.15~eV, the parallel components of electron momentum suffer a broadening of of about 7$\cdot10^{-3}$~\AA$^{-1}$ and the perpendicular component has an average value of about 4$\cdot10^{-3}$~\AA$^{-1}$ and a broadening of 6$\cdot10^{-3}$~\AA$^{-1}$ (we do not show the histograms). The space charge appears to have negligible effects on electron momentum resolution at conditions of photon flux and spot size for which a prominent energy broadening is present. We mention that recent simulations have revealed a mild but appreciable dependence of $\Delta E$ on the emission angle of the photoelectrons \cite{Greco_unpublished}. If we select the electrons with an emission direction close to the axis normal to sample surface, their energy distribution presents a lower broadening with respect to that of the whole photoelectron cloud. Nevertheless these perpendicularly emitted electrons tends to suffer a greater shift towards higher kinetic energies \cite{Greco_unpublished}.

\section{Discussion} \label{sec:discussion}

Our simulations confirmed the pivotal role played by the secondary electrons in the broadening of PES structure also using hard X-rays as exciting radiation. The simple LIK model is effective in providing the order of magnitude of the energy broadening --- though it tends towards an overestimation --- connecting it to the total number $N$ of photoelectrons and the linear dimension $a$ of the photon spot on the sample surface. The  secondary-electron tail is poorly investigated in PES experiments but it includes the most part of photoemitted electrons and it is the main responsible for the loss of resolution in more interesting higher-energy structures. The application of a positive bias to the sample can bring the secondary electrons back to the sample, but it would be ineffective in reducing significantly the energy broadening. The path of the secondary electrons back to the sample cannot be instantaneous while the mutual interaction between photoelectrons causes the energy broadening of the structures within a very short time after the photon pulse. A metallic sample is usually biased in PES experiments applying a voltage between the sample holder and the experimental vacuum chamber. If we consider the studied case of Cu 2p photoemission from copper with $h\nu$=8~keV, we cannot apply a voltage greater than 7~kV, in order to allow the 2p photoelectrons to reach the detector. Because in general sample-holders have dimensions of at least few centimeters, considering the simple model of a spherical capacitor, the electric field on the sample surface $\mathcal{E}$ can hardly exceed a value of $10^6$~V/m. With this electric field, the time that a secondary electron emitted normally to sample surface with an initial energy of 1.6~eV spends out of the sample can be easily calculated as 9~ps. As shown in Fig.~\ref{fig:Cu_2p_LIK}~(a), this time lapse is sufficient to almost complete the broadening of the 2p peaks. Moreover, it is unlikely that such an intense electric field  has no detrimental effect on the flight of higher-energy primary electrons.

If secondary electrons cannot be eliminated, their number can be controlled through the dependence of the quantum yield $\eta$ on the photon energy. We have shown that, increasing the photon energy $h\nu$ from 8000~eV to 9000~eV (crossing the Cu K absorption edge), $\eta$ sharply jumps from $\approx0.004$ to about 0.015, mostly for the larger number of secondary electrons generated by the 1s photoelectrons and by the Auger electrons due to the new recombination channel. From our simulations the energy broadening increases by about a factor four while the number of 2p electrons remains substantially unchanged. A photon energy of 8~keV proves to be a good choice to observe the 2p peaks for the relatively low number of secondary electrons generated. An increase of $h\nu$ above 9~keV would be worthwhile only if we were interested in studying at the same time the 1s photoemission peak. In the design of a pump-and-probe hard X-ray experiment it is then important to choose a radiation energy that minimizes the quantum yield of our sample in relation with the information we want to obtain, in particular avoiding to cross absorption edges whose photoelectrons cannot help our investigation.

A second parameter to work with is obviously the linear dimension of the radiation spot on the sample surface. Nevertheless, a too large spot size makes impossible to resolve spatial variations of composition and electronic properties and can prevent the use of the promising time-of-flight spectrometers. Photoelectrons with the same energy emerging from distant points on the sample surface may have remarkably different times of flight \cite{Lollobrigida_15}, worsening the energy resolution of the apparatus.

It is of fundamental importance to know if the reduced number of impinging photons per pulse, necessary to make tolerable the energy-broadening effect, is sufficient to acquire the PES structures with a satisfactory statistics. This also depends on the repetition rate of the photon source we use for our experiment \cite{Chiang_CT_15}. If we consider the examined $h\nu$=8000~eV photoemission experiment from pure copper, from Fig.~\ref{fig:Cu_2p_LIK}~(b) we expect a tolerable 0.1~eV energy broadening for 2p peaks with $N_{ph}\approx1 \cdot 10^7$ photons per pulse with a moderately large circular spot of radius 0.5~mm. Considering the values of $n_c/N_{ph}$ shown in Table~\ref{tab:Cu_2p} we expect the emission of about 725 2p photoelectrons per photon pulse. Imagining a detector with a reasonable effective angular acceptance of $\pm 2^\circ$ (equivalent to an effective accepted solid angle of 3.8~msr \cite{Lollobrigida_15}) we obtain 0.44 detected 2p photoelectrons per photon pulse. The temporal structure of X-ray beams at the European X-FEL is composed by trains of 2700 pulses separated by 220~ns, with a repetition rate of 10 trains per second \cite{Schneidmiller_techrep_11}. The possibility to exploit 27,000 pulses per second corresponds to a count rate of about 12,000 electrons per second for Cu~2p peaks, which allows a quite good statistics in a reasonable time. Modern TOF analyzers operate as band-pass energy filter, allowing only electrons within a relatively small kinetic-energy window to enter inside the spectrometer and interact with the final detector. Our group has recently presented the design of two electron spectrometers operating in the 5--10~keV region, one based on a cylinder lens and the other on a spherical reflector, with pass-energy windows of 400 and 100~eV, respectively \cite{Lollobrigida_15}. In our example of Cu 2p photoemission with $h\nu$=8000~eV, the band-pass filter would allow to detect in correspondence of every pulse only a number of electrons of the order of unity (Cu 2p electrons and inelastic background electrons with similar energy), avoiding the saturation of the detector due to the huge amount of photoelectrons emitted at the same time.

SASE1 and SASE2 undulators at the European XFEL are expected to produce at a photon energy of 8.27~keV in typical functioning condition (14~GeV electron bunches for a total charge of 0.25~nC and a pulse duration of 23.2~fs) 5.5$\cdot$10$^{11}$ photons per pulse with a FWHM spectrum width $\Delta(h\nu)/h\nu$=0.169\%, corresponding to $\Delta h\nu$=14~eV \cite{Schneidmiller_techrep_11}. A monochromator must be implemented in order to have a radiation on the sample with $\Delta h\nu\approx$0.1~eV, suitable for HAXPES experiments. Considering a very rough proportion, the monochromatization is expected to reduce the radiation flux to about 4$\cdot$10$^9$ photons per pulse \footnote{The energy-time indetermination relation $\Delta (h\nu) \cdot \Delta t \geq \hbar /2$ must be taken into account. For a spectral width $\Delta h\nu$ of about 0.1~eV the pulse duration $\Delta t$ must be larger than $\sim$7~fs.}. A further reduction of this flux by 2 or 3 orders of magnitude is necessary to limit the space-charge broadening to values around tenths of electronvolt.

\section{Conclusions}

In this paper we have discussed how to predict energy broadening of PES structures induced by the space charge with a particular focus on  high-energy time-resolved experiments. We carried out many-body numerical simulations for a cloud of interacting photoelectrons in a HAXPES experiment considering the case of photoemission from metallic Cu. We have confirmed the fundamental role played by the secondary electrons in the broadening of the core-level peaks and verified that this latter is directly proportional to the number $N$ of emitted electrons and inversely proportional to the linear size $a$ of the radiation spot on the sample surface, in accordance with the LIK model. Moreover we have shown that the LIK model is effective in reproducing  the order of magnitude of $\Delta E$ in UV and soft X-ray experiments and it is in acceptable agreement also with our high-energy simulations, demonstrating that it can be used as a rough but simple alternative to many-body calculation. We have found that from the LIK model it is possible to estimate a characteristic time $\tau_C \approx \sqrt{E_0}/a$ which describes the evolution with time of the energy broadening effect due to the interactions in the photoelectron cloud. The space-charge effects on the distribution of the momentum of photoelectrons appear to be unimportant in view of ARPES applications. The energy broadening can be controlled acting on the parameters $N$ and $a$. The number of secondary electrons can be reduced with a convenient choice of the photon energy $h\nu$  that minimizes the quantum yield of the analyzed material preserving the count rate of the core-level photoelectrons. A defocus of the incident beam and the augmented size of the beam spot on the sample surface is the second way to reduce the space-charge effects, but this causes a depletion of spatial resolution. Our simulations on 2p photoemission from polycrystalline Cu have shown that an acceptable energy broadening (0.1~eV or less) is achievable with high statistics (thousands of counts per second) and a reasonable spot dimension (0.5~mm). For the limitations imposed by the space charge, probably PES experiments cannot exploit the full amount of photon flux at disposal at the European XFEL facility but the peculiar time structure of this X-ray source can be tested for innovative time-resolved investigations. The simple case we have illustrated can be used as a general scheme of feasibility study that estimates the maximum usable number of photons per pulse and the corresponding count rate in experiments with various materials.

\section*{Acknowledgements}

The authors wish to thank Dr.~Marco Malvestuto  and Dr.~Martina Dell'Angela for critical reading of the manu\-script and Dr.~Giancarlo Panaccione for fruitful discussions.

This work was partially supported by the ULTRASPIN and EX-PRO-REL Projects in kind (PIK), funded by Ministero dell'Istruzione, dell'Universit\`{a} e della Ricerca (MIUR) through Elettra  Sincrotrone Trieste. G.~G. is grateful to the PIK project EX-PRO-REL for financing her postdoctoral fellowship. V.~L. acknowledges the PIK project ULTRASPIN for granting his fellowship. A.~V. is thankful to Regione Lazio and CRUL for financial support.

\end{document}